\documentclass[letterpaper, 10 pt, conference]{ieeeconf}  % Comment this line out if you need a4paper

\IEEEoverridecommandlockouts                              % This command is only needed if 
\title{\LARGE \bf
PM-EKF: A Physiological Model-Based Extended Kalman Filter for Daily-Life Physical Activity Energy Expenditure Estimation 
}

\author{Shuhao Que$^{1}$, Remco Poelarends$^{2}$, Valentina Breschi$^{3}$, and Ying Wang$^{1}$% <-this % stops a space
\thanks{This study was approved by the ethical review board of University of Twente and funded by the HealthyW8 project.}% <-this % stops a space
\thanks{$^{1}$S. Que and Y. Wang are with the Department of Electrical Engineering, University of Twente, Enschede, The Netherlands.
        {\tt\small s.que@utwente.nl; \tt\small ying.wang@utwente.nl}}%
\thanks{$^{2}$R. Poelarends is with the Department of Nuclear Medicine, Isala, Zwolle, The Netherlands.
        {\tt\small r.j.poelarends@isala.nl}}%
\thanks{$^{3}$V. Breschi is with the Department of Electrical Engineering, Eindhoven University of Technology, Eindhoven, The Netherlands.
        {\tt\small v.breschi@tue.nl}}%
}

\usepackage{url}
\usepackage{graphicx}
\usepackage{float}
\usepackage{amsmath}
\usepackage{amsfonts}

\begin{document}

\maketitle
\thispagestyle{empty}
\pagestyle{empty}

%%%%%%%%%%%%%%%%%%%%%%%%%%%%%%%%%%%%%%%%%%%%%%%%%%%%%%%%%%%%%%%%%%%%
\begin{abstract}
Monitoring physical activity energy expenditure (PAEE) in daily life is essential for characterizing individual health and metabolic status. Although indirect calorimetry provides gold-standard PAEE measurements, it is impractical for continuous daily-life monitoring. Consequently, wearable sensing approaches using inertial measurement units (IMUs) and heart rate (HR) sensors have attracted substantial interest. However, most existing IMU- and HR-based methods are purely data-driven and offer limited physiological interpretability. In this work, we propose a simplified physiological model that explicitly links body movement during activities of daily living to the underlying metabolic gas-exchange processes governing PAEE. The model is formulated as a nonlinear state-space system and embedded within an Extended Kalman Filter (EKF), enabling principled handling of measurement noise, model uncertainty, and system nonlinearities. This physiology-guided EKF filtering framework provides individualized, interpretable PAEE estimates directly from wearable sensors, without employing black-box models. Our model was validated using a dataset, including 9 subjects with around 50 minutes of measurements per subject, collected in our lab simulating a free-living condition. Using the breath-by-breath respiratory data measured by COSMED K5 as reference and explained variance ($R^2$) as evaluation metric, our model's predicted PAEE yielded median (min-max) $R^2$ = 0.72 (0.60--0.87), using three IMUs (pelvis and two thighs) for capturing the body-center-of-mass motion and measured HR for the time-varying cardiac output. Our model outperformed a linear regression (LR) model ($R^2$ = 0.52 (0.23--0.92)) and CNN-LSTM model ($R^2$ = 0.65 (0.46--0.78)) under the same setting on the same dataset. Notably, excluding the sensory HR measurement did not significantly degrade PAEE estimation of all three models, indicating that IMU-captured mechanical workload dominated PAEE estimation performance in our protocol.
\end{abstract}

%%%%%%%%%%%%%%%%%%%%%%%%%%%%%%%%%%%%%%%%%%%%%%%%%%%%%%%%%%%%%%%%%%%%%%%%%%%%%%%%
\section{INTRODUCTION}
Monitoring physical activity energy expenditure (PAEE) in daily life is crucial because it reflects the real-world patterns of human movement and energy balance that underlie health, functional capacity, and disease risk. Specifically, PAEE is a strong and independent predictor of reduced progression toward metabolic syndrome, such as abdominal obesity, hypertension, insulin resistance, and so on~\cite{ekelund2005physical}. Thus, it is of substantial research interest to measure and gain insight into individuals' PAEE. Conventionally, using indirect calorimetry~\cite{b4}, PAEE can be measured with a face mask or a ventilated hood used to collect inhaled air~\cite{b6}. However, such an approach is not sufficiently portable for daily life monitoring~\cite{b9}. As a result, there is an increased focus on the development of wearable sensor-based monitoring tools for PAEE. 

The used wearable sensors usually incorporate inertial measurement units (IMUs), which capture body movement through acceleration measurements~\cite{b11}. IMU-based PAEE estimation approaches range from linear models to complex neural networks (NN), with NN models generally outperforming linear ones~\cite{b11}. Heart rate (HR) is often added as an additional predictor~\cite{b11, b10}, and several studies report improved performance when combining IMU and HR~\cite{santos2014validity, assah2011accuracy}. However, conflicting evidence exists. For example, Mitrzyk et al.~\cite{montoye2018heart} showed that enabling HR-based algorithms did not improve calorie estimation relative to indirect calorimetry in free-living settings.

Despite satisfactory predictive accuracy, most existing approaches rely on data-driven mappings between sensor signals and PAEE without explicitly modeling the physiological processes linking movement (e.g., IMU), cardiovascular response (e.g., HR), and metabolism~\cite{alvarez2020survey}. Specifically, IMU and HR signals function primarily as statistical predictors, without an explicit physical interpretation of how movement constrains metabolic dynamics. Moreover, metabolic information derived from IMU sensors reflects primarily kinematic and mechanical aspects of activity, which capture only part of the total metabolic cost and do not account for additional physiological processes such as internal work, thermoregulation, and individual metabolic efficiency. Consequently, relying solely on IMU-derived proxies may lead to biased and physiologically inconsistent estimates of PAEE, limiting their reliability and generalizability across individuals and activity conditions. This limitation also extends to other existing attempts to enhance explainability by relying on post-hoc analysis or the inclusion of biomechanical features within neural networks~\cite{b12, jung2025estimation}. Although such strategies provide partial explanatory insights, they do not impose explicit physiological constraints. Therefore, they lack intrinsic interpretability beyond explainability, and fail to guarantee causal validity or physiological consistency. As a result, the absence of intrinsic interpretability through mechanistic grounding restricts the ability to meaningfully explain PAEE variations across activities and individuals, limits robustness under distribution shifts, and reduces clinical applicability in contexts where understanding metabolic regulation is essential.

To address these limitations, we propose a Physiological Model-Based Extended Kalman Filter (PM-EKF) approach that integrates a simplified mechanistic PAEE model within an Extended Kalman Filter (EKF)~\cite{ekf2006}. The proposed physiological model captures the essential gas-exchange dynamics governing PAEE during activities of daily living (ADL), while the EKF adaptively fuses model predictions with HR and IMU measurements, which are treated as informative but indirect observations. Specifically, the PAEE is computed from the inferred latent physiological states.

\section{Methods}
The proposed PM-EKF framework consists of three levels, including the signal level, the modeling level, and the inference level. The signal level consists of the preprocessing pipeline of HR and IMU (Section~\ref{sec:preprocessing}) and the metabolic proxy derivation from IMU (Section~\ref{sec:IMUmetabolicproxy}). The modeling level consists of physiological model formulation (Section~\ref{sec:physiomodel}) and state-space formulation (Section~\ref{sec:statespace}). On this level, HR is incorporated as external input driving the physiological process model, whereas IMU-derived metabolic proxies are treated as observations. The inference level includes the extended Kalman filter (EKF) (Section~\ref{sec:ekf}), which estimates latent physiological states that are subsequently used to calculate PAEE (Section~\ref{sec:finalreadout}). The symbols, their units, and corresponding basal or selected values are summarized in the Supplementary Material (“Glossary Table”).

\subsection{Signal level}
\subsubsection{Preprocessing}\label{sec:preprocessing} 
HR is derived from ECG using a modified Pan–Tompkins detector by \cite{thoonen2022movement}. The HR time series is then smoothed with a first-order Savitzky–Golay filter (20~s window), selected empirically for its balance between noise suppression and temporal fidelity, and subsequently resampled to 1~Hz for time alignment with other modalities.

Raw IMU accelerations are corrected for gravity and low-pass filtered using a 4th-order Butterworth filter with a 6~Hz cutoff, consistent with the expected frequency content of ADL. Linear velocity is estimated by integrating the filtered acceleration using the composite trapezoidal rule and resampled to 1~Hz. To counteract drift from residual low-frequency noise, velocity is set to zero whenever the filtered acceleration remains at zero for at least five consecutive samples, following \cite{aranburu2018imu}. The velocity magnitude is then computed to represent the overall velocity of the IMU sensor.

\subsubsection{IMU-derived metabolic proxy}\label{sec:IMUmetabolicproxy}
To obtain the metabolic proxy for the modeling observations in Section~\ref{sec:statespace}, we first estimate the mechanical energy required to move the body from the body velocity, which can be derived from the signals measured from the 3 IMUs (i.e., pelvis and two thighs). Accordingly, the following whole-body kinetic energy equation is used:
\begin{equation} \label{eq:e(t)pelvis_3imu}
    E(t) = \frac{0.5 \cdot (M_U \cdot v^2_{p}(t) + M_L \cdot v^2_{l}(t) + M_R \cdot v^2_{r}(t))}{\mu}
\end{equation}

where $M_L$ and $M_R$ represent the mass of the left and right legs. According to anthropometric segment parameters~\cite{de1996adjustments}, symmetry is assumed such that $M_L = M_R$, each corresponding to approximately 16\% of total body mass when summing the thigh, shank, and foot segments. $M_U$ represents the remaining mass of the whole body. $v_{p}(t)$ denotes the derived velocity from the pelvis IMU, whereas $v_{l}(t)$ and $v_{r}(t)$ denote the derived velocities from the left thigh IMU and the right thigh IMU, respectively. 

$\mu$ denotes the lumped efficiency factor. This dimensionless denominator (0 $<$ $\mu$ $\leq$ 1) is proposed to partially mitigate the fact that IMU-derived segmental kinetic energy, i.e., the nominator of (1), captures only a portion of total physiological energy expenditure due to biomechanical modeling simplifications, unmodeled internal work, and individual movement characteristics. Based on domain knowledge and empirical analysis, we assign 0.06 to all activity types except for cycling, to which we assign 0.02 instead. Because the IMU-derived segmental kinetic energy underestimates the metabolic demand of cycling by an even larger magnitude compared with other activities. Such a choice requires knowing the activity cluster (mode) in advance, a limitation that we plan to overcome in future work, e.g., using multiple-model approaches~\cite[Chapter 11]{bar2001estimation} or an activity recognition algorithm as a pre-step~\cite{que2026synthetic}.

The $E(t)$ is converted to the metabolic consumption rate of oxygen $RM_{O_2}(t)$ using the conversion factor of 19.6 KJ, given that one liter $O_2$ produces 19.6 KJ of heat~\cite{scott2005misconceptions}. This approach assumes that metabolic demand and production approach zero after a period of no body segment movement, consistent with PAEE estimation, which excludes tissue maintenance energy costs. The metabolic production rate of carbon dioxide $RM_{CO_2}(t)$ is estimated using the respiratory quotient $RQ$, which is the ratio of carbon dioxide production to oxygen consumption. Accordingly, we obtain:
\begin{align}
    RM_{O_2}(t) & = \frac{E(t)}{19.6} \label{eq:rmo2} \\
    RM_{CO_2}(t) & = RQ \cdot RM_{O_2}(t) \label{eq:rmco2}
\end{align}

\subsection{Modeling level}
\subsubsection{Physiological model formulation}\label{sec:physiomodel}
\begin{figure}
    \centering
    \includegraphics[width=\linewidth]{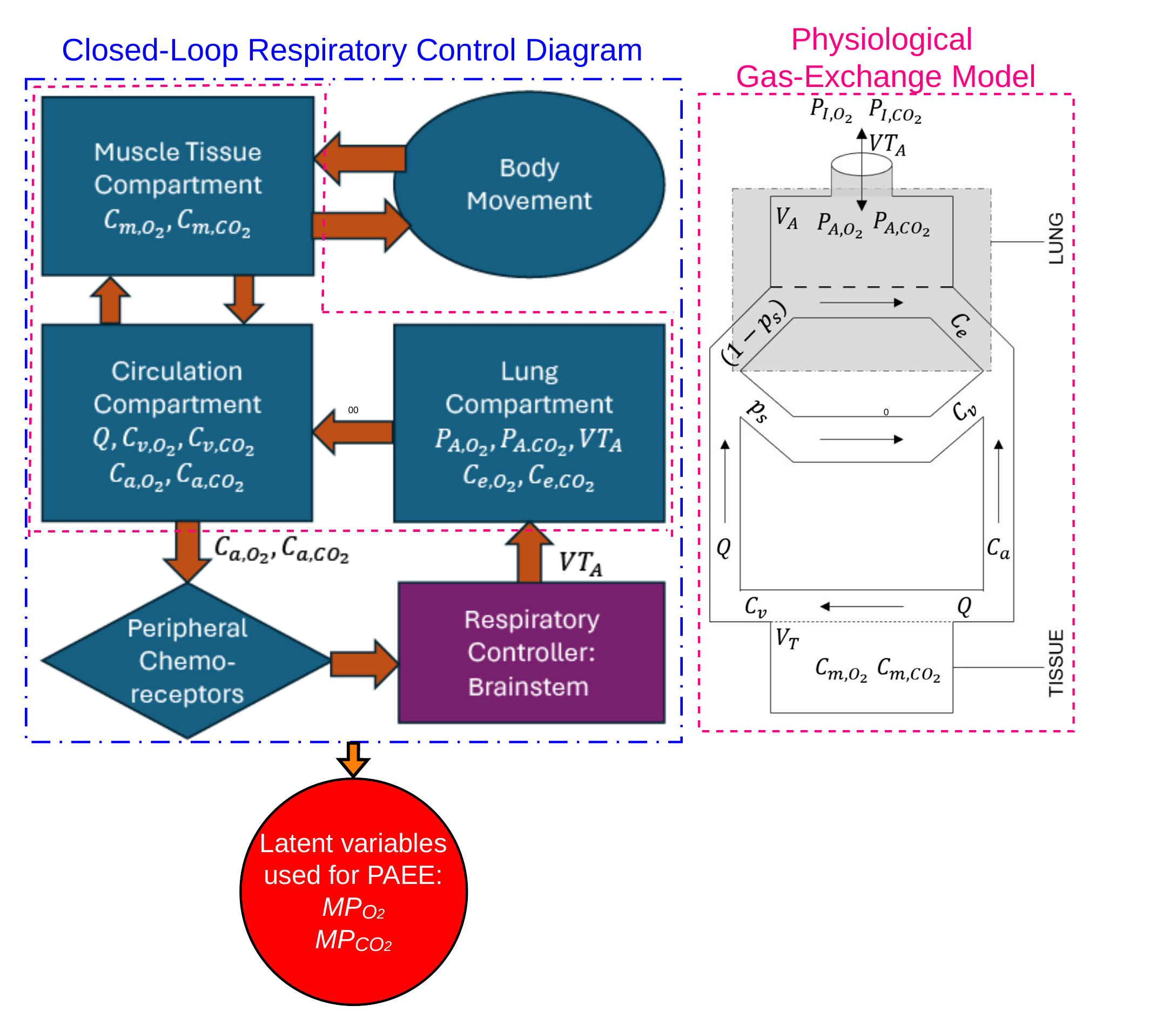}
    \caption{A system-level representation of the respiratory control loop and gas-exchange physiology used to estimate the gas metabolic rates (i.e., $MP_{O_2}$ and $MP_{CO_2}$). These gas metabolic rates are subsequently used to compute physical activity energy expenditure (PAEE) via Weir's formula~\cite{b6}. The physiological gas-exchange model comprises three compartments: the lungs, the circulation, and muscle tissue. This model is regulated by a respiratory controller: the brainstem, which mainly receives input from peripheral chemoreceptors located in the circulation compartment. Heart rate (HR) serves as an external sensory measurement to derive time-varying cardiac output $Q$. Inertial measurement units (IMU) serve as external sensory measurements reflecting body movements during physical activities. $C_{a,O_2}$ and $C_{a,CO_2}$ denote arterial blood concentration of $O_2$ and $CO_2$ in L/L, respectively; $P_{I,O_2}$ and $P_{I,CO_2}$ denote partial pressure of inspired $O_2$ and $CO_2$ in mmHg, respectively; $V_A$ denotes alveolar gas volume in L; $VT_A$ denotes total alveolar ventilation in L/s; $P_{A,O_2}$ and $P_{A, CO_2}$ denote alveolar partial pressure of $O_2$ and $CO_2$ in mmHg, respectively; $p_s$ denotes the dimensionless pulmonary shunt fraction; $C_e$ denotes end-capillary concentration in liters $O_2$ (or $CO_2$) per liter blood; $C_v$ denotes venous concentration in liters $O_2$ (or $CO_2$) per liter blood; $C_a$ denotes arterial concentration in liters $O_2$ (or $CO_2$) per liter blood; $Q$ denotes cardiac output in L/s; $V_T$ denotes tissue volume in L; $C_{v,O_2}$ and $C_{v,CO_2}$ denote venous concentration in liters $O2$ and $CO_2$ per liter blood, respectively. $C_{m,O_2}$ and $C_{m,CO_2}$ represent the effective tissue gas concentrations, in liters $O_2$ (or $CO_2$) per liter blood.}
    \label{fig:model}
\end{figure}

As shown in Figure~\ref{fig:model}, we propose a simplified PAEE physiological model, represented within a respiratory control loop and built around three physiological compartments, i.e., lungs, circulation, and muscle tissue, which together describe the pathways by which oxygen is taken in, transported, and utilized for metabolic energy production underlying PAEE. The model defines a set of latent physiological states that characterize gas exchange and metabolic processes.

\textbf{(a) Model assumptions:}
A total of 15 assumptions are made for our three-compartment physiological model.

The assumptions \textbf{(I)}--\textbf{(V)} are propositions specifically introduced in this work to simplify the model for estimating PAEE from body movement and HR: \textbf{(I)} The PAEE is calculated based on Weir's formula~\cite{b6}, assuming the anaerobic metabolism makes a negligible contribution for the duration of the measurement, which holds for low and moderate intensity activities. \textbf{(II)} The brain compartment used in \cite{chiari1997comprehensive} is omitted, as cerebral metabolism remains relatively constant and contributes minimally to the variations in energy expenditure induced by movement. \textbf{(III)} The respiratory quotient (RQ) is set as a constant value of 0.8, as RQ values typically range between 0.7 during fat-dominant metabolism (e.g., prolonged low-intensity activity) and 1.0 during carbohydrate-dominant metabolism (e.g., high-intensity exercise), with resting mixed-diet values around 0.8~\cite{mcardle2010exercise}. \textbf{(IV)} The $O_2$ stimulus is represented by arterial oxygen concentration $C_{a,O_2}$, while the $CO_2$ stimulus is approximated by alveolar $CO_2$ partial pressure $P_{A,CO_2}$. This approximation is possible because $CO_2$ has a very high diffusive capacity and equilibrates rapidly across the alveolar–capillary membrane. Consequently, arterial and alveolar $CO_2$ tensions remain tightly coupled in healthy individuals, particularly when the pulmonary shunt is small (e.g., $p_s = 0.024$ in this work)~\cite{hardman2003estimating}. \textbf{(V)} The oxygen-carbon dioxide chemoreflex coupling present in \cite{chiari1997comprehensive}, is assumed to be negligible in this work because, in healthy individuals performing ADL, the chemoreflex interactions between hypoxic and hypercapnic stimuli remain weak within the physiological range. Accordingly, the ventilatory response is adequately represented by independent $O_2$ and $CO_2$ control terms without significant loss of fidelity.

\textbf{(VI)} The variables controlled by the respiratory controller are delayed by $T(t)$, partially due to the arterial blood transport, and is inversely proportional to cardiac output~\cite{grodins1967mathematical}. \textbf{(VII)} Following \cite{fincham1983mathematical}, blood pH remains within the physiological range, allowing normal gas-exchange relationships. \textbf{(VIII)} Following \cite{fincham1983mathematical}, dissociation curves are identical for arterial and venous blood. \textbf{(IX)} Following \cite{chiari1997comprehensive}, the lung and muscle tissue volumes (i.e., $V_A$ and $V_T$, respectively) remain constant, as these volumes vary slowly relative to the gas-exchange dynamics and therefore have negligible influence on the short-term kinetics of $O_2$ and $CO_2$ transport. \textbf{(X)} The lungs and muscle tissue are modeled as well-mixed, homogeneous compartments in which pressure and concentration changes occur uniformly and instantaneously, allowing fluctuations in gas exchange to be represented by average compartmental values. \textbf{(XI)} The muscle tissue compartment and venous blood are in rapid equilibrium, i.e., $C_m = C_v$, consistently with \cite{chiari1997comprehensive}. This implies no significant transient gas storage in tissue and allows tissue metabolic rates to be represented as being instantaneously balanced by alveolar gas-exchange fluxes, in accordance with classical lumped-compartment respiratory models. \textbf{(XII)} Blood flow within the circulation compartment is steady and continuous with uniform concentration distribution, as the cardiac output ($Q$) during movement provides an adequate representation of global oxygen delivery and carbon dioxide removal for estimating whole-body PAEE. \textbf{(XIII)} Gas behavior within the lungs adheres to the principles of the ideal gas law, a well-established principle in physics that enables direct relationships between gas partial pressures, concentrations, and alveolar volume. \textbf{(XIV)} Gas exchange across the alveolar–capillary membrane occurs without diffusion limitation, which is valid during low-to-moderate intensity activity, and even during high-intensity activity in healthy individuals, as pulmonary diffusion capacity increases to meet metabolic demand~\cite{turino1963effect}. \textbf{(XV)} Alveolar ventilation ($VT_A$) is controlled by the brainstem only through feedback from peripheral chemoreceptors located in the circulatory system responding to arterial oxygen concentration ($C_{a,O_2}$) and alveolar partial pressure of carbon dioxide ($P_{A,CO_2}$). This hypothesis is consistent with the rapid ventilatory adjustments observed in response to changing metabolic rates during movement~\cite{o1982role, iturriaga2021carotid}. 

\textbf{(b) The muscle tissue and circulation compartments:} Body movement induces changes in the gas concentrations of $O_2$ and $CO_2$ in the muscle tissue, characterized by $C_{m,O_2}$ and $C_{m,CO_2}$. By assuming $C_{m,O_2} = C_{v,O_2}$ and $C_{m,CO_2} = C_{v,CO_2}$ (Assumption (\textbf{XI})), the mass-balance equation for the venous blood/muscle tissue gas concentrations [L/L] is defined as:
\begin{equation} \label{eq:dCvO2}
    V_T \frac{dC_{v,O_2}(t)}{dt} = Q(t) (C_{a,O_2}(t) - C_{v,O_2}(t)) - MP_{O_2}(t)
\end{equation}
\begin{equation} \label{eq:dCvCO2}
\begin{split}
    V_T \frac{dC_{v,CO_2}(t)}{dt} & = Q(t) (C_{a,CO_2}(t) - C_{v,CO_2}(t)) \\
    & + MP_{CO_2}(t)
\end{split}
\end{equation}

where the metabolic production terms $MP_{O_2}$ and $MP_{CO_2}$ represent the muscle's net $O_2$ consumption rate and $CO_2$ production rate during ADL (Section~\ref{sec:finalreadout}). The muscle tissue volume $V_T$ [L] is defined as:
\begin{equation}
    V_T = \frac{M_{SM}}{\rho}
\end{equation}

where $M_{SM}$ denotes the skeletal muscle mass measured by the BIA scale [kg] and $\rho$ denotes the density of skeletal muscle mass [kg/m$^3$].

The venous blood $C_{v,O_2}$ and $C_{v,CO_2}$ flowing through the pulmonary shunt and the blood flowing through the lungs join together to form arterial blood. The arterial gas concentrations [L/L] of $O_2$ and $CO_2$ are calculated by the shunt equation:
\begin{equation} \label{eq:CaO2}
    C_{a,O_2}(t) = (1 - p_s) C_{e,O_2}(t) + p_s C_{v,O_2}(t)
\end{equation}
\begin{equation} \label{eq:CaCO2}
    C_{a,CO_2}(t) = (1 - p_s) C_{e,CO_2}(t) + p_s C_{v,CO_2}(t)
\end{equation}

where $p_s$ represents the pulmonary shunt, giving a constant fraction of venous blood that does not reach the alveoli. 

Following \cite{sullivan1989relation}, the cardiac output $Q(t)$ [L/s] is computed using the following equation:
\begin{equation} \label{eq:Q}
    Q(t) = HR(t)SV(t)
\end{equation}

where $HR(t)$ denotes the measured HR [beat/s] and $SV(t)$ denotes the stroke volume (SV) [L/beat]. Following \cite{zhang2025physiological}, we used the following logarithmic equation to derive the stroke volume from the oxygen uptake $MP_{O_2}(t)$:
\begin{equation}\label{eq:sv}
    SV(t) = \lambda_1 \ln\left(\frac{60 * MP_{O_2}(t)}{1~\text{L/min}}\right) + \lambda_2
\end{equation}

where $\lambda_1$ denotes contractility, which controls how fast SV rises with intensity, and $\lambda_2$ is the baseline stroke volume. From empirical experiments, we selected $\lambda_1$ = 0.02 and $\lambda_2$ = 0.08975 L/beats.

\textbf{(c) The respiratory controller:} The peripheral chemoreceptors detect the changes in arterial blood gas concentrations of $O_2$ and $CO_2$, and send signals to the respiratory controller. Subsequently, the respiratory controller adjusts alveolar ventilation $VT_A$ [L/s] based on deviations from basal levels of $O_2$ and $CO_2$. The regulation of alveolar ventilation $VT_A(t)$ by the respiratory controller is described using a simplified version of the controller presented in \cite{chiari1997comprehensive}:
\begin{equation} \label{eq:dvdotA}
    \begin{split}
        \tau \frac{dVT_A(t)}{dt} & = -G_{p, O_2} C_{a, O_2} (t-T(t)) + \\
        & G_{p, CO_2} P_{A, CO_2}(t-T(t)) + K_1 - VT_A(t)
    \end{split}
\end{equation}

where $\tau$ is the dynamic time constant, here set to 1 second, while $G_{p, O_2}$ and $G_{p, CO_2}$ represent the proportional gains in the controller. The parameter $K_1$ is defined so that $VT_A$ = 0 L/s in basal conditions, i.e.,
\begin{equation}
    K_1 = G_{p,O_2}C_{a,O_2,0} - G_{p,CO_2}P_{a,CO_2,0}
\end{equation}

where $C_{a,O_2,0}$ [L/L] and $P_{a,CO_2,0}$ [mmHg] are the basal values of $C_{a,O_2}$ and $P_{a,CO_2}$, respectively.

To prevent ventilation $VT_A$ from reaching negative values, as this is physiologically unattainable, ventilation is automatically set to zero whenever its value falls below zero. 

According to (\ref{eq:dvdotA}), all controlled variables are delayed by $T(t)$ (Assumption (\textbf{VI})), defined as:
\begin{equation}
    T(t) = \frac{K_T}{Q(t)}
\end{equation}

where $K_T$ is a constant parameter, whose value is chosen to provide a delay of 6 seconds when $Q(t)$ is at its basal level~\cite{chiari1997comprehensive}.

\textbf{(d) The lung compartment:} The alveolar ventilation $VT_A(t)$ adjusted by the respiratory controller acts as the input to the pulmonary gas-exchange dynamics.

In the lung compartment, the balance equations are used to describe the conservation of mass for $O_2$ and $CO_2$. Still leveraging \cite{chiari1997comprehensive}, the following dynamic equations of alveolar partial pressure [mmHg] are used:
\begin{equation}\label{eq:dPaO2}
    \begin{split}
        V_A \frac{dP_{A,O_2}(t)}{dt}  & = VT_A(t) (P_{I,O_2} - P_{A,O_2}(t)) + 
    \\
    & \lambda Q(t) (1 - p_s) (C_{v,O_2}(t) - C_{e,O_2}(t))
    \end{split}
\end{equation}
\begin{equation}\label{eq:dPaCO2}
    \begin{split}
        V_A \frac{dP_{A,CO_2}(t)}{dt}  & = VT_A(t) (P_{I,CO_2} - P_{A,CO_2}(t)) + 
    \\
    & \lambda Q(t) (1 - p_s) (C_{v,CO_2}(t) - C_{e,CO_2}(t))
    \end{split}
\end{equation}

where $V_A$ [L] denotes the alveolar gas volume and $P_{I,O_2}$ [mmHg] and $P_{I,CO_2}$ [mmHg] denote the partial pressure of $O_2$ and $CO_2$ in the inspired gas, respectively. $C_{e,O_2}(t)$ and $C_{e,CO_2}(t)$ denote the end-capillary blood concentrations of $O_2$ and $CO_2$ [L/L at Standard Temperature and Pressure, Dry (STPD)], respectively. $\lambda$ is a constant coefficient that converts blood concentrations at STPD into alveolar partial pressure at Body Temperature and Pressure, Saturated (BTPS)~\cite{chiari1997comprehensive}.

Assuming rapid diffusion and near-equilibrium at the alveolar–capillary interface (Assumption \textbf{(XIV)}), alveolar pressure is taken to be equal to end-capillary pressure. Dissociation curves from \cite{fincham1983mathematical} are used to convert alveolar partial pressures to end-capillary concentrations:
\begin{align}
    C_{e, O_2}(t) &= K_2  \left(1 - e^{-K_3 P_{A, O_2}(t)}\right)^2 \label{eq:ceo2} \\
    C_{e, CO_2}(t) &= K_4  P_{A, CO_2}(t) \label{eq:ceco2}
\end{align}

where $K_2$, $K_3$, and $K_4$ represent constants determined through a linear best fit~\cite{fincham1983mathematical}.

\subsubsection{State-space formulation}\label{sec:statespace}
In this subsection, we convert our proposed physiological model into a state-space format, comprising a process model and a measurement model.

\textbf{(a) Process model:} We compactly denote the process model:
\begin{equation}
    \Dot{\mathbf{x}}(t) = f(\mathbf{x}(t), u(t))
\end{equation}

where $u(t)$ is the known external input in beats per second (bps), i.e., 
\begin{equation}
    u(t) = \frac{HR(t)}{60}
\end{equation}

According to the model described in Section~\ref{sec:physiomodel}, the state vector is defined as:
\begin{equation}\label{eq:statevector}
    \mathbf{x}(t) = \begin{bmatrix}
        x_1(t) \\
        x_2(t) \\
        x_3(t) \\
        x_4(t) \\
        x_5(t)
    \end{bmatrix} = \begin{bmatrix}
        P_{A,O_2}(t) \\
        P_{A,CO_2}(t) \\
        C_{v,O_2}(t) \\
        C_{v,CO_2}(t) \\
        VT_A(t)
    \end{bmatrix}
\end{equation}

To avoid repetition with existing equations presented in Section~\ref{sec:physiomodel}, the detailed formulation of the state dynamics $\mathbf{\dot{x}}(t)$ is provided in the Supplementary Material (“State Vector Dynamics”).

\textbf{(b) Measurement model:} The observation function $h(\mathbf{x}(t), u(t))$ represents the pulmonary gas-exchange fluxes and is derived from the three-compartment physiological gas-exchange formulation~\cite{chiari1997comprehensive}. It maps the latent physiological states and cardiac activity to instantaneous oxygen consumption and carbon dioxide production rates. The IMU-derived metabolic proxies, $RM_{O_2}$ and $RM_{CO_2}$, serve as the measurement $\mathbf{z}(t)$, providing an external observation of movement-related metabolic demand.

Accordingly, the observation function is defined as\footnote{For readability, the time dependence (t) of state variable and external input is omitted.}:
\begin{equation}
        h(\mathbf{x}, u)  = \begin{bmatrix}
            h_1(\mathbf{x}, u) \\
            h_2(\mathbf{x}, u)
        \end{bmatrix}
\end{equation}

where
\begin{subequations}\label{eq:measurement_model}
\begin{align}
h_1(\mathbf{x}, u) & = uv(x_1,x_5)(1-p_s)(K_2\left(1 - e^{-K_3x_1}\right)^2 - x_3)
\label{eq:obser_1} \\
h_2(\mathbf{x}, u) & = uv(x_1,x_5)(1-p_s)(K_4x_2 - x_4)
\label{eq:obser_2} 
\end{align}
\end{subequations}

Subsequently, the measurement model is defined as:
\begin{equation}
    \mathbf{z}(t) = h(\mathbf{x}(t), u(t))
\end{equation}

\subsection{Inference level}
\subsubsection{Extended Kalman filter}\label{sec:ekf}
On the inference level, the EKF framework is utilized to recursively estimate the states in (\ref{eq:statevector}) over time. At each time step, the process model propagates the physiological states according to the gas-exchange dynamics driven by cardiac activity, while the EKF corrects the process model's state prediction using the IMU-derived metabolic proxies $RM_{O_2}$ and $RM_{CO_2}$. As the EKF procedure is standard, the detailed formulation is provided in the Supplementary Material (“Extended Kalman Filter”).

\subsubsection{Physiological readout}\label{sec:finalreadout}
After the hidden states are inferred using EKF, the PAEE is calculated based on indirect calorimetry, which leverages the exhaled gas volumes by alveolar ventilation.

The exhaled gas volumes by alveolar ventilation can be estimated as the product of exhaled volume and the difference in gas fraction of inhaled and exhaled air, for $O_2$ and $CO_2$. The gas fraction of exhaled air is estimated using the alveolar gas equation~\cite{wijayasiri2017primary}, which is substituted in the following equations:
\begin{align}
    VT_{O_2}(t) & = VT_A(t) (F_{I, O_2} - \frac{P_{A, O_2}(t)}{P_{atm} - P_{H_2O}}) \label{eq:vo2} \\
    VT_{CO_2}(t) & = - VT_A(t) (F_{I, CO_2} - \frac{P_{A, CO_2}(t)}{P_{atm} - P_{H_2O}}) \label{eq:vco2}
\end{align}

where $VT_{O_2}(t)$ and $VT_{CO_2}(t)$ denote the exhaled volumes of $O_2$ and $CO_2$ per second [L/s], respectively. $F_{I, O_2}$ and $F_{I, CO_2}$ are the partial gas fractions of inhaled air, $P_{atm}$ [mmHg] denotes the atmospheric pressure at sea level, and $P_{H_2O}$ [mmHg] denotes the partial pressure of water in the alveoli, which is used to correct for the humidity levels in the alveoli.

Accordingly, the gas metabolic production and elimination rate [L/s] are defined as:
\begin{equation}\label{eq:MPO2}
    \begin{split}
        MP_{O_2}(t) & = B\cdot VT_{O_2}(t) 
    \end{split}
\end{equation}
\begin{equation}\label{eq:MPCO2}
    \begin{split}
        MP_{CO_2}(t) & = B\cdot VT_{CO_2}(t)
    \end{split}
\end{equation}

where the factor $B$ converts the gas volume from BTPS to STPD~\cite{pittman2016} for gas metabolic rate calculation~\cite{b6}, which is defined as:
\begin{equation}
    B = \frac{P_{atm} - P_{H_2O}}{P_{atm}}\cdot \frac{273}{310}
\end{equation}

where 273~K is the standard temperature in Kelvin ($\approx$ 0~$^\circ$C) used in STPD, and 310~K is the body temperature in Kelvin ($\approx$ 37$^\circ$C) used in BTPS.

Based on indirect calorimetry, the gas metabolic rate of $O_2$ and $CO_2$ per second are used to estimate PAEE [kcal/s] during ADL using Weir's formula~\cite{b6}:
\begin{equation} \label{eq:paee}
    E_{PA}(t) = 3.9 \cdot MP_{O_2}(t) + 1.1 \cdot MP_{CO_2}(t)
\end{equation}

\subsection{Data collection}
The proposed PM-EKF methodology is validated on a dataset collected at the eHealth House, University of Twente (doi: 10.5281/zenodo.19388844). A total of 10 participants (30\% female) were recruited for this study. Inclusion criteria were: (1) aged between 18 and 60~\cite{pontzer2021daily}; (2) have a Body Mass Index (BMI) lower than 40 kg/m2~\cite{westerterp2017control}; (3) free of cardiovascular diseases, respiratory diseases, metabolic disorders; (4) not being pregnant for female participants; (5) free of physical disabilities that impact daily living. The study was ethically approved by the Ethics Committee of Computer \& Information Science of the University of Twente (File no. 230728). Informed consent was obtained from all individual participants included in the study. Static data, including age, sex, height, and weight, were collected. Body composition was estimated using a Bioelectrical Impedance Analysis (BIA) scale (Omron BF511).

\begin{table}[h]
\caption{List of activities with their corresponding duration and intensity. NF (not fixed) represents variable activity duration. *Despite the activity being considered moderate-to-high intensity, they remain submaximal and predominantly aerobic in the present healthy cohort.}
\label{tab:activities}
\begin{center}
\begin{tabular}{|c||c||c|}
\hline
Intensity & Activity & Duration [s]\\
\hline
Low & Sitting resting & 300 \\
Low & Sitting reading & 300 \\
Low & Standing still & 180 \\
Low & Surveying (on a laptop) & NF \\
Low & Emptying dishwasher & NF \\
Moderate & Mopping & NF \\
Moderate & Stacking shelves with books & NF \\
Moderate & Treadmill (3 km/h) & 300 \\
Moderate & Treadmill (5 km/h) & 300 \\
Moderate--High* & Climbing stairs (5 times) & NF \\
Moderate--High* & Cycle at 125 Watt & 300 \\
\hline
\end{tabular}
\end{center}
\end{table}

\begin{figure}
    \centering
    \includegraphics[width=\linewidth]{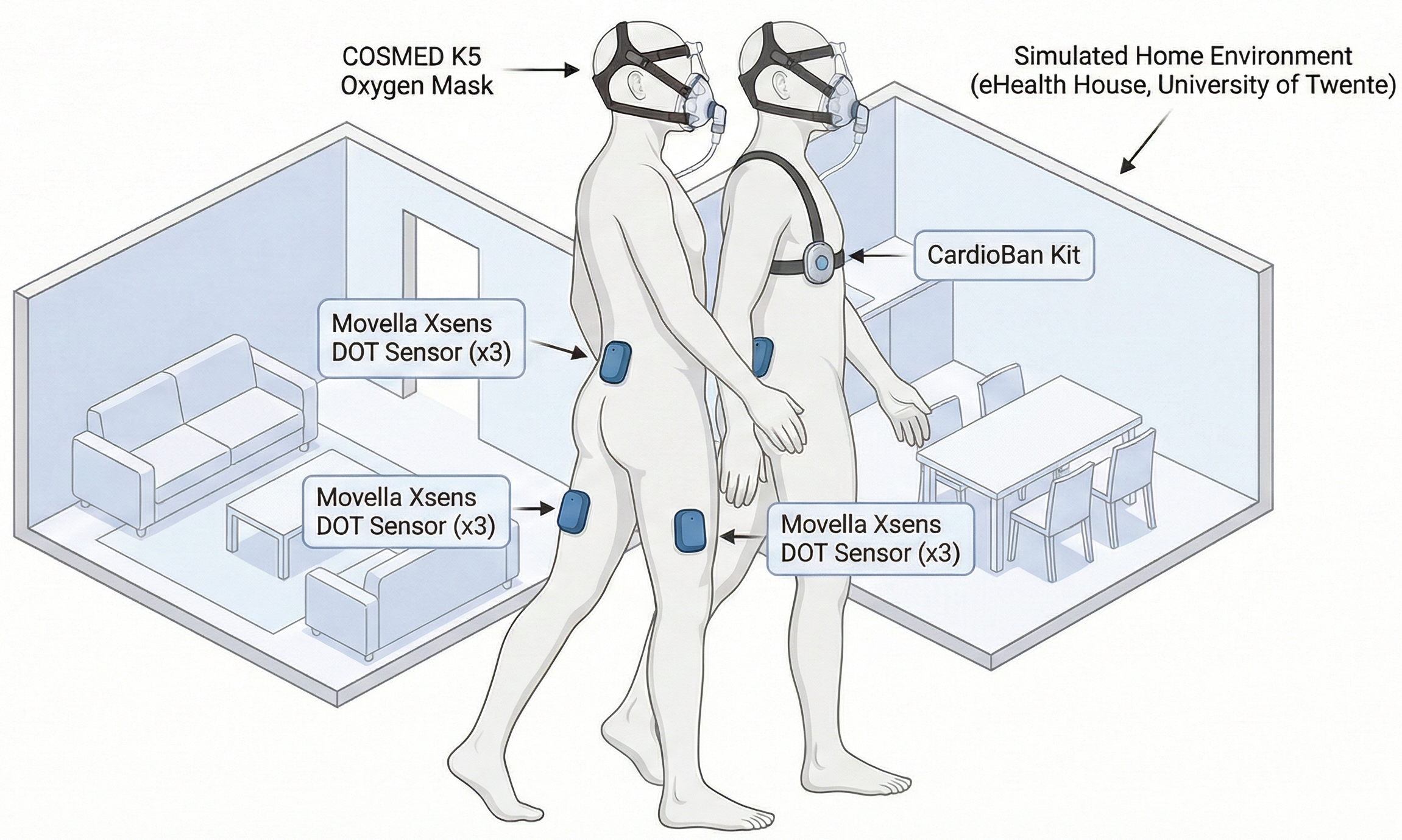}
    \caption{Placement of the used sensors, including three Movella Xsens DOT sensors for the inertial measurements, a CardioBan Kit for electrocardiogram measurements, and COSMED K5 for the breath-by-breath respiratory data.}
    \label{fig:sensors}
\end{figure}

Each data collection session commenced with a 30-minute quiet rest in the supine position to estimate rest metabolic rate (RMR)~\cite{nosslinger2021underestimation}. A series of ADL followed this, as presented in Table \ref{tab:activities}. The activities were categorized with respect to their metabolic intensity following the guidelines of \cite{ainsworth2000compendium}. Although climbing stairs and cycling were categorized as moderate-to-high intensity, they remained submaximal and predominantly aerobic within the current study cohort. Hence, the assumptions underlying Weir’s formula hold. Most activities were performed for at least 5 minutes to reach steady-state energy expenditure, as recommended by \cite{alvarez2020survey}. For emptying the dishwasher, mopping, stacking shelves, climbing stairs, and surveying, the participants were instructed to execute the activities at their own pace, which made the duration of these activities variable. Each participant performed all activities in a randomized order to prevent the introduction of bias in the dataset~\cite{paraschiakos2022recurrent}. The randomization of the performed activity order and the variable duration of certain activities, as mentioned before, help to simulate real situations of daily living.

IMU data was collected at 30 Hz at three body locations, i.e., left thigh, right thigh, and pelvis, using Movella Xsens DOT~\cite{movelladot}. Single-lead Electrocardiogram (ECG) data was collected using CardioBan Kit~\cite{cardioban} at 80 Hz. Breath-by-breath respiratory data, serving as ground truth, was collected using COSMED K5~\cite{deblois2021reliability}. Activity labels were manually annotated by the researchers based on video recordings from five fixed cameras installed throughout the eHealth House using the OMNIA COSMED software. The placement of the aforementioned wearable sensors is shown in Figure \ref{fig:sensors}.

To estimate the RMR from $O_2$ consumption and $CO_2$ production, the first 5 minutes of the RMR recording were discarded to remove transient effects following the onset of rest. The mean volumes of $O_2$ and $CO_2$ computed from the remaining period were taken as the participant’s RMR, following the recommendations of \cite{compher2006best}. Since total energy expenditure (TEE) during ADL comprises both RMR and PAEE, the RMR-contributed $O_2$ and $CO_2$ volumes were subtracted from TEE to obtain activity-related gas-exchange signals. For multi-modal signal synchronization, the breath-by-breath $O_2$ and $CO_2$ volume data were resampled to 1 Hz and smoothed using a first-order Savitzky–Golay filter (20 s window) to reduce noise and mitigate artifacts from talking.

\subsection{Model evaluation}
The proposed PM-EKF model is evaluated through a two-fold complementary scheme: In the first fold, we examine the structural soundness of the proposed PM-EKF formulation. This includes (1) observability analysis, where we assess the observability of the physiological state-space formulation from IMU and HR measurements; and (2) a physiological interpretability and plausibility evaluation of the inferred latent physiological states and the PAEE computed from them, evaluating their realism and consistency with known metabolic dynamics. In the second fold, we perform an external comparison with two purely data-driven approaches, namely a linear regression (LR) model and a convolutional neural network–long short-term memory (CNN-LSTM) model. This comparison includes (1) overall performance evaluation of PAEE estimation accuracy, and (2) per-activity-intensity performance evaluation. (3) We further assess each model’s dependence on HR input by excluding HR from LR and CNN-LSTM, and by fixing HR to a constant value (70 bpm) in the PM-EKF model. This analysis directly addresses the IMU-only versus IMU+HR debate, as stated in the Introduction.

\subsubsection{Observability analysis}
The observability of the latent physiological states of the proposed PM-EKF framework is assessed from two aspects: the overall observability and state estimation uncertainty.

\textbf{(a) Overall observability:} As part of the model-level validation, we conduct an overall observability analysis to assess whether the latent physiological states can, in principle, be uniquely inferred from the respiratory metabolic measurements $RM_{O_2}$ and $RM_{CO_2}$. Since our state-space model is nonlinear, we analyze its observability using the Lie derivative, from which the Hermann-Krener observability matrix~\cite{hermann2003nonlinear}, $\mathcal{O}_{HK} \in \mathbb{R}^{N_o \times n_x}$, along a nominal trajectory $\mathbf{x}_0(t), u_0(t)$ can be constructed. A full rank of $\mathcal{O}_{HK}$ indicates that all states are theoretically observable. To quantify the contribution of each individual state to the overall observability, we compute the column-wise 2-norm of $\mathcal{O}_{HK}$, which yields a per-state observability score, denoted as $s_i$. The normalized score $\tilde{s}_i$ ranging from 0 to 1 is then used, where higher values indicate stronger observability. The detailed derivations are provided in the Supplementary Material ("Observability Analysis").

\textbf{(b) State estimation uncertainty:} The uncertainty of the state estimation can be quantified by the confidence interval along the state prediction trajectory derived from the EKF state covariance $\mathbf{P}_{k|k}$ (see the Supplementary Material ("Extended Kalman Filter")). It provides a principled and time-resolved quantification of state estimation reliability, revealing how uncertainty dynamically adapts to activity-dependent observability and measurement informativeness.

\subsubsection{Physiological interpretability and plausibility evaluation}
The physiological plausibility is evaluated on two aspects: the physiological interpretability of the PM-EKF model and the plausibility of the predicted PAEE values from both the PM-EKF model and the benchmark models defined in Section~\ref{sec:baselinemodels}.

\textbf{(a) Interpretability evaluation:} 
\begin{table}[H]
\caption{Reference physiological bounds used to assess PM-EKF’s interpretability across intensity categories, as presented in Table~\ref{tab:activities}. Bounds are approximate ranges for healthy adults at sea level under submaximal aerobic conditions~\cite{guyton2006medical}.}
\label{tab:state_bounds}
\centering
\begin{tabular}{|c||c||c||c|}
\hline
\textbf{State (unit)} &
\textbf{Low} &
\textbf{Moderate} &
\textbf{Moderate--High} \\
 & & & (submaximal)\\
\hline
$P_{A,O_2}$ (mmHg) &
$90$--$110$ &
$95$--$120$ &
$100$--$130$ \\
\hline
$P_{A,CO_2}$ (mmHg) &
$38$--$45$ &
$34$--$42$ &
$30$--$38$ \\
\hline
$C_{v,O_2}$ (L/L) &
$0.13$--$0.16$ &
$0.10$--$0.14$ &
$0.07$--$0.12$ \\
\hline
$C_{v,CO_2}$ (L/L) &
$0.58$--$0.64$ &
$0.62$--$0.70$ &
$0.65$--$0.75$ \\
\hline
$VT_A$ (L/s) &
$0.04$--$0.12$ &
$0.15$--$0.40$ &
$0.30$--$0.80$ \\
\hline
\end{tabular}
\end{table}

The interpretability is assessed by examining whether qualitatively the temporal evolution and quantitatively magnitudes of the inferred states are consistent with established respiratory and metabolic physiology~\cite{guyton2006medical}. Because the latent states are not measured using non-invasive devices but computationally derived by the physiological model, there is no ground-truth reference for individual states. As a result, a numerical comparison between estimated and true state trajectories is not feasible. For quantitative magnitude analysis, the physiological bounds for each state variable during low, moderate, and moderate-high intensity activities are provided in Table~\ref{tab:state_bounds}. 

\textbf{(b) PAEE prediction plausibility:} To validate the physiological plausibility of the predicted PAEE from all models, a physiological PAEE lower bound 0~kcal/s is used. The violation rate can thus be derived as the percentage of the total number of predicted PAEE values lower than 0 kcal/s over that of the PAEE ground-truth measurements from all subjects

\subsubsection{Performance evaluation}\label{sec:baselinemodels}
Following the same 3-IMU and HR setting, we use CNN-LSTM and LR models as baselines to compare with our PM-EKF model.

We adopt the CNN–LSTM proposed by \cite{lee2024imu}, which ingests 6-axis IMU sequences (accelerometer and gyroscope) and does not incorporate HR measurements. To fit the scope of the present study, several modifications are introduced to the original CNN–LSTM architecture. (1) First, only 30-s windows of triaxial accelerometer data are used as sequential input, and gyroscope signals are omitted. (2) Second, HR is incorporated as an auxiliary physiological input by using the instantaneous HR value at the 30th second of each HR time series window and concatenating it with the LSTM-derived feature representation prior to the final fully-connected (FC) layer. This ensures that the injected HR retains its instantaneous nature rather than representing a window-averaged value, consistent with how HR is incorporated in the EKF framework. (3) Third, all input signals are normalized on a per-participant basis, and sequences are zero-padded at the window boundaries to ensure uniform input length. Finally, the FC layer maps the combined representation to the predicted PAEE. The same hyperparameters from \cite{lee2024imu} are used for the model architecture and training.

For LR, we use 3 IMUs based on our previous work~\cite{que2025accelerometry}, where the feature, namely the total integrated absolute acceleration (IAA), is calculated based on \cite{bouten1994assessment}. To incorporate HR, we include the instantaneous HR value defined above as the additional regressor.

\textbf{(a) Overall performance:} The coefficient of determination (i.e., explained variance):
\begin{equation}
R^2 = 1 - \frac{\sum_{i=1}^n (E_{PA,i} - \hat{E}_{PA,i})^2}{\sum_{i=1}^n (E_{PA,i} - \bar{E}_{PA})^2}
\end{equation}

is used to evaluate the model's predicted PAEE against the ground-truth measurements. $E_{PA,i}$ is the true PAEE values, $\hat{E}_{PA,i}$ is the predicted PAEE values, $\bar{E}_{PA}$ is the mean of the true values, and $n$ is the number of data points. 

Since the CNN–LSTM and LR models are purely data-driven, their performance is evaluated and compared using leave-one-subject-out cross-validation. The PM-EKF model is evaluated on the same left-out subject in each cross-validation fold, ensuring a consistent evaluation and comparison protocol across all models.

\textbf{(b) Per-activity-intensity performance:} For per-activity-intensity analysis, all models are first fitted on the full measurement set. The resulting PAEE estimates are then segmented according to activity, and per-activity mean-normalized root-mean-squared error (NRMSE) is computed separately for each activity segment. The per-activity NRMSE scores are then grouped together (according to Table~\ref{tab:activities}) to derive median values as per-activity-intensity NRMSE. NRMSE is used instead of $R^2$ because it provides an absolute, scale-normalized error measure, whereas $R^2$ reflects relative variance explained and can be unstable or less informative for short, low-variance activity segments. The NRMSE between the ground-truth and predicted PAEE over one activity segment is define as:
\begin{equation}
    NRMSE = \frac{\sqrt{\sum^n_{i=1} \frac{(E_{PA,i} - \hat{E}_{PA,i})^2}{n}}}{\bar{E}_{PA}}
\end{equation}

where $E_{PA,i}$ is the true PAEE value, $\hat{E}_{PA,i}$ is the predicted PAEE value, $\bar{E}_{PA}$ is the mean of the true PAEE values, and $n$ is the number of data points. 

An NRMSE value of 1 indicates performance equivalent to a mean predictor, i.e., predicting the mean of the observed values. 

\textbf{(c) HR-dependence analysis:} To assess the overall dependence (i.e., based on $R^2$) of all models on HR measurements, HR input is removed. For the PM-EKF, the measured HR signal is replaced with a fixed value of 70 bpm throughout the time sequence. For the LR and CNN-LSTM models, HR is excluded entirely during both training and testing, following the implementation strategy described in our previous work~\cite{que2025accelerometry}.

\textbf{(d) Statistical test:} The non-parametric Wilcoxon signed-rank paired test~\cite{ref_statisticaltest} is used to evaluate if there is a statistically significant difference among the PAEE predicted with our PM-EKF model, the LR model, and the CNN-LSTM model. It is also used to evaluate the performance of all models with and without HR sensory measurement. The same paired test is also used to perform within-model pairwise comparisons across predictions over different activities for each model. In these comparisons, the Bonferroni correction~\cite{bonferroni1936teoria} is applied to mitigate the bias introduced by multiple-pair comparisons.

\section{Results}
\subsection{Observability analysis}
\subsubsection{Overall observability}
According to the Lie-derivative along the EKF-predicted state trajectory, our PM-EKF model maintains a full rank of 5 for the observability matrix $\mathcal{O}_{HK}$, indicating that all five state variables $P_{A,O_2}, P_{A,CO_2}, C_{v,O_2}, C_{v,CO_2}, VT_A$ are locally observable at the linearization point along the predicted state trajectory.

Although a full rank points to the theoretical observability of all states, the mean per-state observability quantifies their practical observability. States $P_{A,CO_2}$ and $C_{v,CO_2}$ yield mean observability scores of 0.99 and 0.90, respectively, indicating strong structural observability under the proposed measurement configuration. This pattern is consistent with the observation function, in which $CO_2$-related concentration gradients directly modulate the predicted pulmonary gas-exchange flux. States $P_{A,O_2}$ and $C_{v,O_2}$ yield mean observability scores of 0.18 and 0.19, respectively, suggesting reduced sensitivity within the observation mapping. This is consistent with the fact that $O_2$-related variables exhibit comparatively smaller dynamic variation in the gas-exchange formulation under submaximal daily-life activities (Assumptions \textbf{(I)} and \textbf{(XIV)}), which reduces their effective observability within the proposed measurement configuration. The state $VT_A$ yield a mean observability score of 0.06, indicating weak instantaneous sensitivity within the measurement Jacobian. Although ventilation physiologically governs gas exchange, its effect enters the observation model primarily through dynamical coupling in the process equations rather than directly through the measurement mapping. Consequently, its reconstruction relies mainly on indirect dynamical interactions with more observable states.

\subsubsection{State estimation uncertainty}
Similar time-evolving pattern of state estimation uncertainty during the entire measurement is observed across all subjects. The confidence intervals derived from the EKF state covariance $\mathbf{P}_{k|k}$ vary coherently with activity intensity. They narrow during moderate and moderate-to-high intensity activities, where model predictions and measurements strongly constrain the states. The confidence intervals widen during low-intensity activities, where observability is lower, and measurement variability is higher. This collective pattern indicates that the EKF is functioning consistently across states, adapting its uncertainty in accordance with the informativeness of available measurements.

\subsection{Physiological interpretability and plausibility analysis}
\subsubsection{PM-EKF interpretability}
The interpretability of the PM-EKF is observed from all subjects. According to Table~\ref{tab:state_bounds}, all five state variables from all subjects remain within the expected submaximal aerobic range throughout the protocol. A state evolution analysis summarised from all subjects is provided below.

During low intensity activities, $P_{A,O_2}$ stabilizes at the lower end of its range and $P_{A,CO_2}$ at the higher end, while $C_{v,O_2}$ remains relatively high and $C_{v,CO_2}$ low, reflecting reduced $O_2$ extraction, lower $CO_2$ production, and diminished ventilatory drive, consistent with low $VT_A$. At moderate to moderate-to-high intensities, $P_{A,O_2}$ increases and $P_{A,CO_2}$ decreases, while $C_{v,O_2}$ decreases and $C_{v,CO_2}$ increases, indicating elevated $O_2$ uptake, increased $CO_2$ clearance, and greater metabolic demand, accompanied by increased $VT_A$. Across activity transitions, all states evolve smoothly, consistent with the continuous dynamics of gas exchange, venous return, and respiratory control.

\subsubsection{PAEE prediction plausibility}
The physiological violation rates for CNN-LSTM, LR, and PM-EKF are 0.53\%, 0.03\%, 0\%, respectively. This shows that PM-EKF strictly abides by the physiological lower bound in terms of predicting PAEE, whereas CNN-LSTM and LR occasionally violate such a principle, especially CNN-LSTM, which is a highly data-driven function approximator implicitly capturing data dependence.

\subsection{Performance analysis}
\subsubsection{Overall performance}
\begin{figure*}
    \centering
    \includegraphics[width=\linewidth]{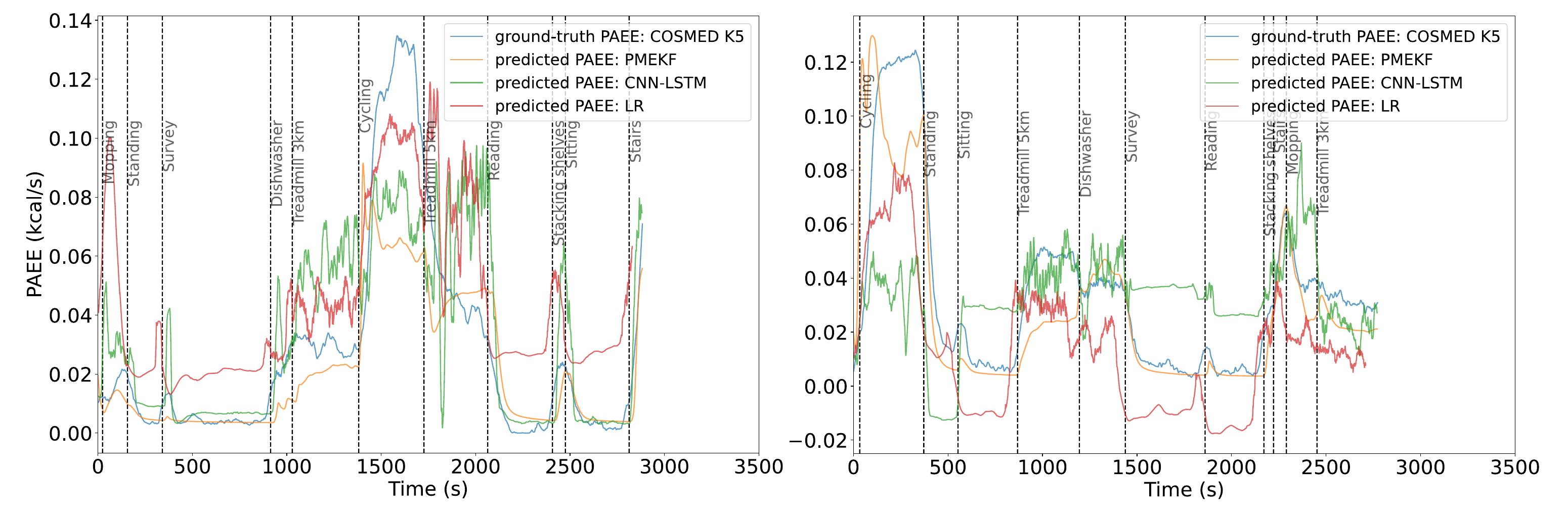}
    \caption{PAEE estimation results using the PM-EKF, CNN-LSTM, and LR models on two subjects in the time domain. The left and right figures correspond to when the PM-EKF performs the worst and the best, respectively.}
    \label{fig:eesubjs47}
\end{figure*}

In Table~\ref{tab:R2withHR}, regarding the overall performance with HR, the PM-EKF model yields significantly better performance than CNN-LSTM (p-value = 0.039) and LR (p-value = 0.012). Example prediction results of all three models in the time domain are presented in Figure~\ref{fig:eesubjs47}. The left and right subjects are selected as cases where the PM-EKF performs the worst and the best, respectively. In the case where the PM-EKF performs the worst, it can be observed that the PM-EKF underestimates the cycling PAEE by the largest magnitude compared with LR and CNN-LSTM. In the other case, where the PM-EKF performs the best, the PM-EKF model captures well the general PAEE trend across different activity types. In contrast, both the LR and CNN-LSTM underestimate the cycling PAEE by larger magnitudes in this case. In addition, for standing, both LR and CNN-LSTM models have produced physiologically implausible PAEE values (i.e., $<$ 0). For other low-tensity activities, both models predict PAEE values excessively higher than the reference values. Per-subject performance based on $R^2$ of each model is provided in the Supplementary Material ("Per-Subject Performance").

\begin{table}[H]
\caption{Results of physical activity energy expenditure estimation with and without heart rate using explained variance ($R^2$) and mean-normalized root-mean-squared error (NRMSE). The highest achieved median $R^2$ and lowest median NRMSE values for each row are marked in bold font.}
\label{tab:R2withHR}
\begin{center}
\resizebox{0.95\linewidth}{!}{%
\begin{tabular}{|cc|cc|cc|cc|}
\hline
\multicolumn{1}{|c}{} &
\multicolumn{1}{c|}{} &
\multicolumn{2}{c|}{PM-EKF} &
\multicolumn{2}{c|}{CNN-LSTM} &
\multicolumn{2}{c|}{LR} \\
\hline
 & & HR & No HR & HR & No HR & HR & No HR\\
\hline
$R^2$ & & & & & & & \\
Overall & Median & 0.72 & \textbf{0.76} & 0.65 & 0.55 & 0.52 & 0.48\\
& Min--Max & 0.60--0.87 & 0.63--0.81 & 0.46--0.78 & -0.06--0.78 & 0.23--0.92 & -0.00--0.86\\
\hline
$NRMSE$ & & & & & & & \\
Activity Intensity & & & & & & & \\
Low & Median & 0.19 & \textbf{0.19} & 1.05 & 0.88 & 1.36 & 0.79 \\
& Min--Max & 0.05--0.64 & 0.05--0.66 & 0.15--9.23 & 0.26--6.40 & 0.13--27.56 & 0.27--39.77 \\
Moderate & Median & 0.45 & 0.45 & \textbf{0.32} & 0.39 & 0.54 & 0.46 \\
& Min--Max & 0.20--0.89 & 0.22--0.78 & 0.11--2.82 & 0.11--1.16 & 0.08--3.27 & 0.15--1.43\\
Moderate--High & Median & 0.67 & 0.69 & \textbf{0.29} & 0.35 & 0.37 & 0.34 \\
& Min--Max & 0.14--1.87 & 0.12--1.76 & 0.15--0.50 & 0.24--0.71 & 0.09--1.21 & 0.16--0.78\\
\hline
\end{tabular}
}
\end{center}
\end{table}

\subsubsection{Per-activity-intensity performance}
According to the within-model pairwise p-value analysis of the NRMSE (see Table~\ref{tab:R2withHR}) results, the PM-EKF model exhibits statistically consistent performance across activities of low and moderate intensities. For the moderate-high intensity, the PM-EKF model performs significantly worse. The CNN-LSTM model performs the best for the moderate-high intensity, while the worst for the low intensity. For the moderate intensity activities, although the CNN-LSTM model yields a good median NRMSE value, the NRMSE range is large (i.e., 2.71), especially in comparison with PM-EKF (NRMSE range: 0.69), indicating a relatively large performance inconsistency across different subjects. Similarly to CNN-LSTM, the LR model yields the worst results for the low intensity and the best for the moderate-high intensity. The same performance inconsistency is also observed for the moderate intensity activities, with a relatively large NRMSE range (i.e., 3.15).

\subsubsection{HR-dependence analysis}
It can be observed in Table~\ref{tab:R2withHR} that our PM-EKF model with and without HR measurements yield similar performance with no statistically significant difference. The same applies to the LR model. The CNN-LSTM model exhibits a performance decline numerically. However, according to the paired test, such a decline is not statistically significant, with a p-value of 0.129. We speculate that this is due to the limited data size (9 subjects). Without HR sensory measurements, the PM-EKF model outperforms the CNN-LSTM (p-value = 0.004) and LR (p-value = 0.008).

\section{Discussions}
Among the three models, the PM-EKF achieves significantly higher overall accuracy (median $R^2$ = 0.76) than the CNN-LSTM and LR models, shows physiological interpretability, and has the best physiological plausibility, with a 0\% violation rate. It performs best at low intensity (NRMSE = 0.19), while showing slightly higher errors at higher intensities. In contrast, CNN-LSTM and LR perform better at moderate and moderate–high intensities but degrade markedly at low intensity (NRMSE $>$ 1, i.e., worse than a mean predictor) and can yield physiologically implausible negative PAEE estimates. No statistically significant differences are observed between conditions with and without HR for any model.

The PM-EKF maintains consistent performance across most activities, with a notable exception in cycling, where PAEE is systematically underestimated. This limitation arises from the reduced observability of cycling-specific mechanical work when using IMU-derived metabolic proxies ($RM_{O_2}$ and $RM_{CO_2}$), which cannot adequately capture external power contributions (e.g., ergometer resistance). Nevertheless, the model remains physiologically plausible and interpretable, as its gas-exchange foundation constrains state evolution and allows this underestimation to be explained mechanistically. In contrast, CNN-LSTM and LR achieve relatively better performance during cycling but degrade substantially at low intensities. This bias suggests that, without embedded physiological structure, these models preferentially fit higher-variance, higher-magnitude patterns, at the expense of predictive accuracy in low-intensity regimes. Additionally, the cycling limitation of the PM-EKF is partly due to the use of a fixed efficiency factor $\mu$ across subjects and different activities. Incorporating $\mu$ as an augmented EKF state in future work could enable real-time, subject-and-activity-specific adaptation, potentially improving both accuracy and physiological fidelity.

Besides the potentially limited sample size, the absence of a significant contribution of HR to PAEE prediction across the three models may be attributed to the predominantly low-to-moderate intensity ADL in our dataset. During such activities, energy expenditure is largely governed by the metabolic cost of mechanical work, while HR reflects cardiovascular and autonomic regulation rather than directly determining energy turnover~\cite{rennie2001estimating}. Moreover, the included moderate-to-high intensity activities (i.e., cycling and climbing stairs) involve substantial lower-limb motion, which is well captured by thigh-mounted IMUs, further reducing the relative contribution of HR. In contrast, for moderate-to-high activities with limited lower-limb movement (e.g., resistance training), HR may play a more important role in PAEE estimation. 

The observability analysis provides a mechanistic explanation for the physiological consistency of the PM-EKF. Although the $O_2$-related states and $VT_A$ exhibit relatively low practical observability, the PM-EKF model maintains robust PAEE estimation due to its physiological structure. Through gas-exchange dynamics, $O_2$ and $CO_2$ states are coupled via mass-balance relationships and shared ventilatory processes, allowing highly observable $CO_2$-related states to indirectly constrain less observable states. Moreover, PAEE is derived from integrated metabolic fluxes rather than from any single instantaneous state variable. Consequently, even when certain states exhibit weak instantaneous sensitivity in the measurement mapping, their trajectories remain dynamically constrained within physiologically plausible bounds, enabling stable and accurate PAEE reconstruction.

Another key factor underlying the physiological consistency of the PM-EKF is the role of IMU signals. Unlike CNN-LSTM and LR, the PM-EKF does not directly regress IMU signals to PAEE. Instead, IMU-derived metabolic proxies based on Newtonian mechanics are integrated within a physiologically constrained state-space framework. This structured integration enforces physically meaningful relationships, leading to more consistent and plausible estimates across activities.

Clinically, the PM-EKF enables individualized PAEE monitoring without prior model training. Wearable IMU sensory measurements are integrated as observations, while the EKF dynamically reconciles model predictions with noisy measurements, allowing the system to adapt to subject-specific physiology and varying conditions. This joint modeling approach provides physiological interpretability, plausibility, and robust PAEE estimates in daily-life settings. In addition, the inferred latent physiological states offer potential for longitudinal monitoring and may serve as interpretable biomarkers.

The general limitations of this work include the relatively limited sample size and the lack of explicit consideration of demographic factors such as sex, age, and BMI. The limited dataset may restrict the statistical power and generalizability of the findings, particularly across diverse populations and activity patterns. In addition, demographic characteristics are known to influence physiological responses and energy expenditure, and their omission may introduce bias or reduce the model’s ability to capture inter-individual variability.

\section{Conclusions}
This work introduces a methodological framework for subject-specific PAEE estimation from wearable IMU and HR sensors. In addition to accurate PAEE estimation, the framework also provides interpretable latent physiological states with potential clinical value. By embedding physiological structure within a state-space and EKF inference framework, the approach enables personalized and physiologically consistent estimation without data-driven training. In addition, the insignificant contribution of HR shows that IMU measurements alone are sufficient to capture movement-driven metabolic dynamics within the ADL included in our protocol. Future work will focus on further validating and improving the PM-EKF framework on a larger and more diverse (e.g., more ADL types) dataset.

%\addtolength{\textheight}{-12cm}   % This command serves to balance the column lengths
                                  % on the last page of the document manually. It shortens
                                  % the textheight of the last page by a suitable amount.
                                  % This command does not take effect until the next page
                                  % so it should come on the page before the last. Make
                                  % sure that you do not shorten the textheight too much.

%%%%%%%%%%%%%%%%%%%%%%%%%%%%%%%%%%%%%%%%%%%%%%%%%%%%%%%%%%%%%%%%%%%%%%%%%%%%%%%%

%%%%%%%%%%%%%%%%%%%%%%%%%%%%%%%%%%%%%%%%%%%%%%%%%%%%%%%%%%%%%%%%%%%%%%%%%%%%%%%%

%%%%%%%%%%%%%%%%%%%%%%%%%%%%%%%%%%%%%%%%%%%%%%%%%%%%%%%%%%%%%%%%%%%%%%%%%%%%%%%%

\section*{ACKNOWLEDGMENT}
We acknowledge Prof. Peter Veltink for his contributions in providing initial feedback on the physiological modeling part. Funded by the European Union (the HealthyW8 project). Views and opinions expressed are however those of the authors only and do not necessarily reflect those of the European Union or HaDEA. Neither the European Union nor the granting authority can be held responsible for them.

%%%%%%%%%%%%%%%%%%%%%%%%%%%%%%%%%%%%%%%%%%%%%%%%%%%%%%%%%%%%%%%%%%%%%%%%%%%%%%%%

\end{document}